\documentclass[aps,prb,showpacs,amsmath,amssymb,reprint,superscriptaddress]{revtex4-1}
\pdfoutput=1

\usepackage{graphicx}% Include figure files
\usepackage{color}
\usepackage{amsmath}
\usepackage{amssymb}
\usepackage{amsfonts}
\usepackage{hyperref}
\usepackage{bm}
\usepackage{xcolor}
\usepackage{soul}

\newcommand{\dd}[1]{\mathrm{d}#1\,}

\DeclareMathOperator{\tr}{tr}

\definecolor{PV-color}{rgb}{0.97,0.57,0.11}

\definecolor{FA-color}{rgb}{0.0,0.57,0.11}

\definecolor{TTH-color}{rgb}{0.0,0.0,1}

\begin{document}

%\preprint{}

\title{Superconductivity near a magnetic domain wall
}


\author{Faluke Aikebaier}
\email[]{faluke.aikebaier@jyu.fi}
\affiliation{Department of Physics and Nanoscience Center, University of Jyv\"{a}skyl\"{a}, P.O. Box 35 (YFL), FI-40014 University of Jyv\"{a}skyl\"{a}, Finland
}

\author{P. Virtanen}
\affiliation{NEST, Istituto Nanoscienze-CNR and Scuola Normale Superiore, I-56127 Pisa, Italy}

\author{Tero T. Heikkil\"{a}}
\affiliation{Department of Physics and Nanoscience Center, University of Jyv\"{a}skyl\"{a}, P.O. Box 35 (YFL), FI-40014 University of Jyv\"{a}skyl\"{a}, Finland
}
%\thanks{}
%\altaffiliation{University of Jyväskylä}

%\email[]{}
%\thanks{}
%\altaffiliation{University of Jyväskylä}


\date{\today}

\begin{abstract}
We study the equilibrium properties of a ferromagnetic insulator/superconductor structure near a magnetic domain wall. We show how the domain wall size is affected by the superconductivity in such structures. Moreover, we calculate several physical quantities altered due to the magnetic domain wall, such as the spin current density and local density of states, as well as the resulting tunneling conductance into a structure with a magnetic domain wall.
\end{abstract}


\maketitle

\section{Introduction\label{sec:introduction}}
%\section{\label{sec:introduction}}
%\subsection{}
%\subsubsection{}
One of the hallmarks of superconductivity is the Meissner effect, where a superconductor expels an external magnetic field from its interior.\cite{Meissner1933} Increasing the magnetic field, superconductivity gets suppressed and eventually destroyed by fields exceeding a critical field. However, the critical field of thin films in parallel magnetic field is greatly increased than that of the bulk superconductors, since the Meissner effect is negligibly small in this case.\cite{tinkham2004introduction} Rather, the critical field is determined by the paramagnetic effect.\cite{chandrasekhar1962note,clogston1962upper} This effect suppresses superconductivity by aligning the electrons of Cooper pairs to be separated in energy. The magnetic field in this case then leads to a Zeeman effect.  

One important consequence of the Zeeman field is the splitting of the density of states (DoS) in energy.\cite{PhysRevLett.25.1270,PhysRevLett.26.192,PhysRevB.11.4224,PhysRevB.16.4907,PhysRevB.22.1331} In the absence of the Zeeman field, the DoS of a superconductor shows a singularity at the energy which is equal to the superconducting pair potential. This singularity is separated by the Zeeman field for each spin species. The spin splitting can also be induced by making contact of a superconductor (S) with a ferromagnetic insulator (FI).\cite{PhysRevLett.56.1746,PhysRevB.42.8235,0953-8984-19-16-165202,PhysRevLett.106.247001,senapati2011spin,PhysRevB.90.144509} In this case, the ferromagnetic insulator induces a strong exchange field with a small external magnetic field or even in its absence.

Various properties of superconductors with a spin-splitting density of states have been studied in recent years. An example is the strong thermoelectric effect with a thermopower predicted to exceed $k_B/e$ and a possibility of obtaining large values of the figure of merit $ZT \gg 1$ at low temperatures\cite{PhysRevLett.110.047002,1367-2630-16-7-073002,PhysRevLett.112.057001}. Indications of this effect were also recently  detected\cite{PhysRevLett.116.097001,PhysRevB.95.224505}. A spin accumulation in such a structure was detected at much longer distances than the relaxation lengths at the normal state.\cite{quay2013spin,PhysRevLett.109.207001,PhysRevB.87.024517,PhysRevLett.114.167002} Such effects are reviewed in Refs.~\onlinecite{RevModPhys.90.041001,2017arXiv170608245B}.

In the above studies, the spin-splitting field induced in the superconductors is considered to be homogeneous. However, this is not always the case in a FI/S structure due to the nonuniformity of the ferromagnets.  The nonuniformity can be described by a domain structure. Since the typical domain size in ferromagnets is much longer than the superconducting coherence length, the uniform magnetization is a good assumption in many cases. Here we relax this assumption and study the effects of inhomogeneity, especially domain walls.

The inhomogeneous exchange field also brings interesting effects to superconductivity. The existence of spin-singlet superconductivity in a magnetic domain wall in a ferromagnet (F) was studied several decades ago.\cite{buzdin1984existence} This effect also has been studied in S/F bilayers,\cite{aladyshkin2003domain,houzet2006theory} where a superconducting material is placed on top of a ferromagnet with a domain structure. Decreasing the temperature, superconductivity first appears just above the domain wall. The experimental realization of such domain wall superconductivity was reported in Ref.~\onlinecite{yang2004domain}. The generation of a spin supercurrent in Josephson contacts with a domain wall is discussed in Ref.~\onlinecite{Shomali_2011}. The reconstructed density of states at the end of the superconductor\cite{2018arXiv181104373B} and the magnetoelectric effects\cite{2018arXiv181109304R} in S/F bilayers with a magnetic texture were also studied recently. In the case of FI/S structures, a peculiar tunneling conductance was observed in a recent experiment and the magnetic domains in the ferromagnet was considered responsible for the experimental result.\cite{strambini2017revealing}

Usually, the inhomogeneity of the exchange field induced by the ferromagnetic insulator in the superconductor can be represented by a multi-domain structure, namely, alternating domains with opposite magnetization directions connected via domain walls. The theoretical model of the tunneling conductance in Ref.~\onlinecite{strambini2017revealing} concentrates on this case. Since the size of the domains is often much longer than the superconducting coherence length, whereas here we consider a single domain wall structure separating two domains with opposite magnetization directions. Moreover, the size of the domain wall in Ref.~\onlinecite{strambini2017revealing} was considered much smaller than the superconducting coherence length, here we also consider larger domain walls and study several physical quantities altered due to the inhomogeneity, such as the spin current density and the local density of states.

This paper is organized as follows. We introduce the FI/S structure containing a magnetic domain wall in Sec.~\ref{sec:ModelandMethod}, and solve the Usadel equation with an extended $\theta$-parameterization. Equilibrium spin current density, as a direct consequence of the inhomogeneous exchange field, and its various properties are discussed in Sec.~\ref{sec:EquilibriumSpinCurrentDensity}. In Sec.~\ref{sec:Effectofsuperconductivityonthedomainwallsize} we discuss the effect of superconductivity on the domain wall size, by considering the contribution of the superconducting free energy to the domain wall energy, and show that the maximal relative effect happens when the domain wall size is of the order of $\xi_0$, the superconducting coherence length (here defined at zero temperature and exchange field). For the possible measurement accessing this physics, we discuss the local density of states in Sec.~\ref{sec:Densityofstates} and the tunneling conductance across a nearby tunneling barrier in Sec.~\ref{sec:Tunnelingcurrentandconductance}. 

\section{Model and Method\label{sec:ModelandMethod}}
%\section{\label{sec:ModelandMethod}}
\subsection{Model}
%\subsubsection{}

\begin{figure}[t]
\includegraphics[width=1.0\linewidth]{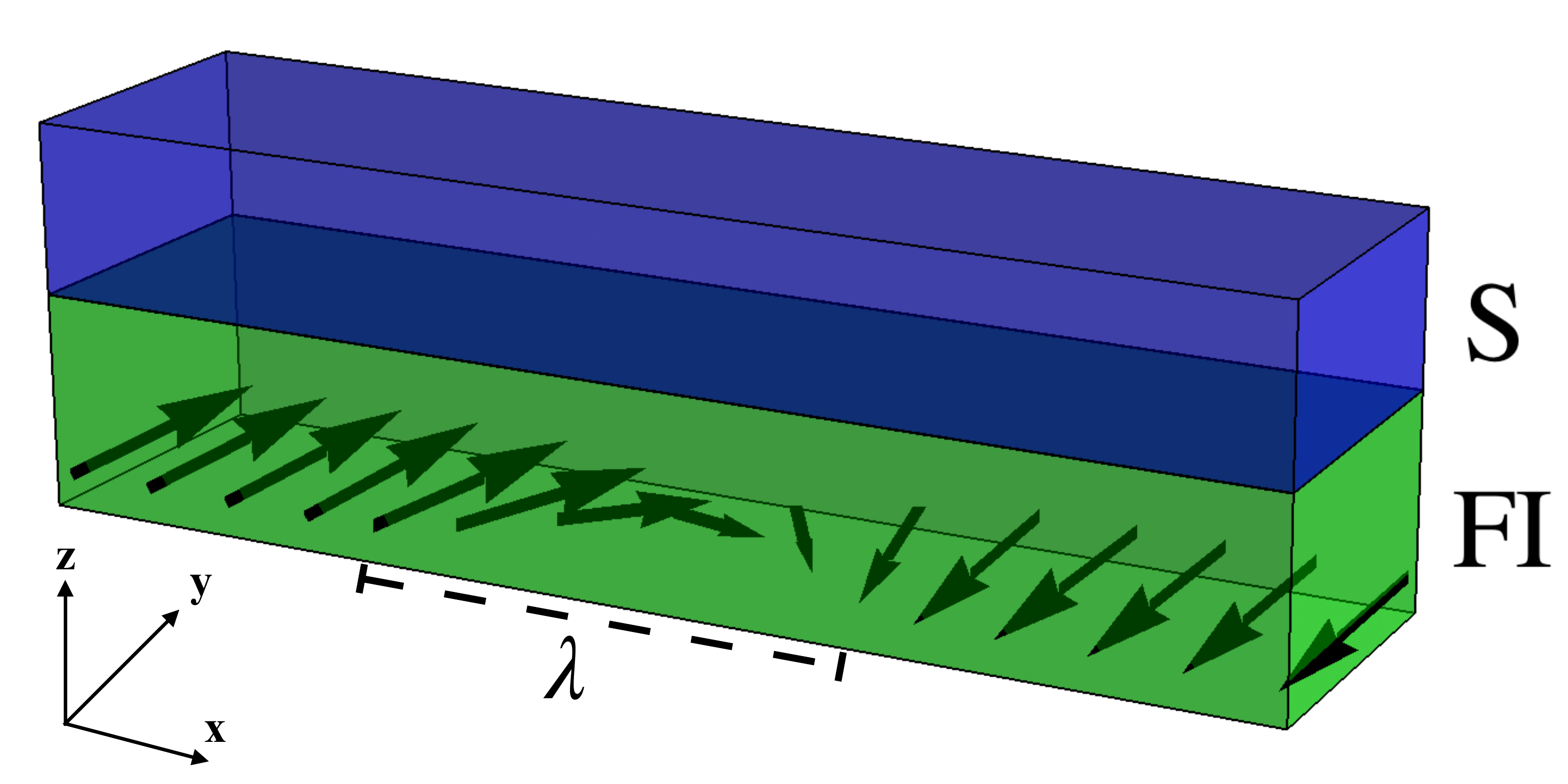}% Here is how to import EPS art
\caption{\label{fig:model} Schematic view of the stucture considered in this paper. A superconducting wire is placed on top of a FI wire containing a magnetic domain wall. The center of the domain wall is located at $x=0$.}
\end{figure}

We study the properties of the structure in Fig.~\ref{fig:model} in equilibrium. A superconducting wire is placed on top of a FI wire containing a magnetic domain wall. In many thin ferromagnets the domain wall structure energetically prefers a N\'eel type, in which the rotation of the magnetization happens in the plane of the domain wall. Therefore we consider a N\'eel domain wall with size $\lambda$. We make a variational ansatz and define the magnetization rotation angle as
\begin{equation}\label{eq:RotationAngle}
\begin{split}
\alpha(x)=&\frac{\pi}{\lambda}\left(x+\frac{\lambda}{2}\right)\Theta\left(x+\frac{\lambda}{2}\right)\Theta\left(\frac{\lambda}{2}-x\right)\\
&+\pi\Theta\left(x-\frac{\lambda}{2}\right),
\end{split}
\end{equation}
where $\Theta$ is the heaviside step function. The choice of $\alpha(x)$ in Eq.~\eqref{eq:RotationAngle} as a linear function of $x$ instead of the typically used hyperbolic functions that lead to a somewhat lower energy brings certain technical advantages. Its derivative is a constant inside the domain wall. This simplifies the Usadel equation, which describes the properties of the superconductor (see Sec.~\ref{subsec:RotationMatrix}). The non-analytic derivative of $\alpha(x)$ at the boundary of the domain wall, can be transferred to the boundary conditions of the Usadel equation (see Sec.~\ref{subsec:Boundarycondition}). Besides, the domain wall energy introduced by this choice of $\alpha(x)$ gives only slightly larger energy than the rotation angle constructed with hyperbolic functions (see Sec.~\ref{sec:Effectofsuperconductivityonthedomainwallsize}). 

The properties of a superconductor in the diffusive limit, namely, in the case when the elastic relaxation rate is much larger than any other energy scales in the problem, are described by the Usadel equation. In equilibrium, it is convenient to use the Usadel equation in the Matsubara representation ($\hbar=k_B=1$)
\begin{equation}\label{eq:UsadelEquation}
D\nabla\cdot\left(\check{g}\nabla\check{g}\right)-
\left[\omega_n\hat{\tau}_3+i\boldsymbol{h}\cdot\boldsymbol{\sigma}\hat{\tau}_3+\check{\Delta}+\check{\Sigma},\check{g}\right]=0,
\end{equation}
where $D$ is the diffusion constant, and $\check{g}$ is the quasiclassical Green's function satisfying the normalization condition $\check{g}^2=\check{1}$. In the commutator, $\omega_n=(2n+1)\pi T$ is the Matsubara frequency, $T$ is the temperature and $n$ is an integer, $\boldsymbol{h}$ is the exchange field, $\boldsymbol{\sigma}=(\hat{\sigma}_1,\hat{\sigma}_2,\hat{\sigma}_3)$ is a vector of Pauli spin matrices, and $\check{\Delta}=\Delta\hat{\tau}_1$ is the superconducting pair potential. The self-energy $\check{\Sigma}=\check{\Sigma}_{so}+\check{\Sigma}_{sf}$, where $\check{\Sigma}_{so}=\boldsymbol{\sigma}\cdot\check{g}\boldsymbol{\sigma}/(8\tau_{so})$ and $\check{\Sigma}_{sf}=\boldsymbol{\sigma}\cdot\hat{\tau}_3\check{g}\hat{\tau}_3\boldsymbol{\sigma}/(8\tau_{sf})$ describe spin and charge imbalance relaxation due to the spin-orbit scattering and exchange interaction with magnetic impurities with corresponding relaxation times $\tau_{so}$ and $\tau_{sf}$, and the Pauli matrices $\hat{\tau}_j(\hat{\sigma}_j)$ are in the Nambu (spin) space. We choose the Nambu spinor as
$$
\Psi=\begin{pmatrix}
\psi_{\uparrow}(x)& \psi_{\downarrow}(x)& -\psi_{\downarrow}^{\dagger}(x)& \psi_{\uparrow}^{\dagger}(x)
\end{pmatrix}^{\top},
$$
where $\top$ denotes a transpose. 

With the rotation angle described in Eq.~\eqref{eq:RotationAngle}, the exchange field can be written as
$$
\boldsymbol{h}=h(\sin\alpha(x),0,\cos\alpha(x)),
$$
where $h$ is the exchange field strength. With this choice, the exchange field depends only on $x$, and rotates in the $xz$-plane, namely, $h_y=0$ everywhere. Correspondingly, the gradient in Eq.~\eqref{eq:UsadelEquation} becomes a derivative in the $x$-direction.

The Usadel equation we apply here is based on a lowest-order spherical harmonics expansion of the Green's function in terms of the momentum direction. Hence it cannot describe the domain wall superconductivity in a d-wave superconductor/ferromagnetic insulator multilayers.~\cite{PhysRevLett.121.077003} The physics studied in Ref.~\onlinecite{PhysRevLett.121.077003} is hence outside the scope of the present work.

\subsection{Rotation matrix}
\label{subsec:RotationMatrix}
By introducing a position dependent rotation matrix, we can rotate the spin axis parallel to the local magnetization direction, so that the inhomogeneous exchange field in Eq.~\eqref{eq:UsadelEquation} can be treated as homogeneous. We define the rotation matrix as
$$
\check{R}=e^{i\hat{\sigma}_2\alpha(x)/2},
$$
where $\alpha(x)$ is the rotation angle in Eq.~\eqref{eq:RotationAngle}, and $\hat{\sigma}_y$ is the second Pauli matrix. Considering this rotation matrix, we define a new quasiclassical Green's function $\check{g}_0$
\begin{equation}\label{eq:FullGreensFunction}
\check{g}=\check{R}^{\dagger}\check{g}_0\check{R},
\end{equation}
so that $\check{g}_0$ satisfies
\begin{equation}\label{eq:RotUsaEqu}
D\check{\partial}_x^A\cdot\left(\check{g}_0\check{\partial}_x^A\check{g}_0 \right)-\left[\omega_n\hat{\tau}_3+ih\hat{\sigma}_3\hat{\tau}_3+\check{\Delta}+\check{\Sigma},\check{g}_0 \right]=0,
\end{equation}
where 
\begin{equation}\label{eq:LongDerivative}
\check{\partial}_x^AX=\partial_xX-\left[ A,X\right],
\end{equation}
$$
A=i\hat{\sigma}_2\alpha'(x)/2.
$$
Thus the problem reduces to solving the Green's function $\check{g}_0$ for a homogeneous exchange field $h$, but with a redefined gradient with an SU(2) type vector potential $A$. Moreover, it is also straightforward to show that $\check{g}_0$ satisfies the normalization condition $\check{g}_0^2=\check{g}^2=\check{1}$.

The long derivative in Eq.~\eqref{eq:RotUsaEqu} with the form in Eq.~\eqref{eq:LongDerivative} introduces some extra terms with respect to the usual derivative and commutator terms in the Usadel equation. One of them is of the form
\begin{equation}\label{eq:SpiOrbRelLikTer}
D\left[A,\check{g}_0A\check{g}_0 \right]=-\frac{1}{2}\alpha'(x)^2D\left[\hat{\sigma}_2\check{g}_0\hat{\sigma}_2,\check{g}_0 \right].
\end{equation}
This term has a similar form as the spin-orbit relaxation with a relaxation rate $\alpha'(x)^2D/2$, but only in one spin direction. It is hence similar to the intrinsic (Rashba or Dresselhaus) spin-orbit coupling.\cite{PhysRevLett.110.117003,PhysRevB.89.134517,PhysRevB.82.195316} Since $\alpha'(x)\neq0$ inside the domain wall, this term is a direct outcome of the existence of the inhomogeneous exchange field. In other words, inhomogeneous exchange field acts like a spin-orbit relaxation, reducing the effect of the exchange field without affecting the superconducting pair potential. This is shown on some physical quantities in the following sections. 

\subsection{Boundary condition}
\label{subsec:Boundarycondition}
The long derivative $\partial^A_x g$ in the Usadel equation in Eq.~\eqref{eq:RotUsaEqu} has to be continuous, so that a discontinuity in the derivative of $\alpha(x)$ implies a discontinuity in the derivative of $\check{g}$ at $x=\pm\lambda/2$. In order to describe this discontinuity, we integrate Eq.~\eqref{eq:RotUsaEqu} at the boundary, obtaining
\begin{equation}\label{eq:BoundCond}
\check{g}_0\partial_x\check{g}_0\biggr\vert_{x_b^\pm}-\check{g}_0\partial_x\check{g}_0\biggr\vert_{x_b^\mp}=\frac{1}{2}\alpha'(x)\left[\check{g}_0i\hat{\sigma}_2,\check{g}_0\right]\biggr\vert_{x_b^\pm},
\end{equation}
where $x_b=\pm\lambda/2$ and $\pm$ refers to the right and left side of the boundary. These boundary conditions together with the solutions in the case of a homogeneous exchange field far from the domain wall form the boundary conditions to the solutions of the Usadel equation in Eq.~\eqref{eq:RotUsaEqu}.

\subsection{Parameterization}
The quasiclassical Green's function $\check{g}$ and $\check{g}_0$ are $4\times4$ matrices in the Nambu $\otimes$ spin space. Since $\check{g}$ represents the Green's function for the inhomogeneous exchange field, we parameterize $\check{g}_0$ following the parameterization of the quasiclassical Green's function for inhomogeneous exchange field in Ref.~\onlinecite{ivanov2006minigap},
\begin{equation}\label{eq:ParamOfg0}
\begin{split}
\check{g}_0=&\cos\theta\hat{\tau}_3\left(M_0+i\tan\theta\boldsymbol{M}\cdot\boldsymbol{\sigma} \right)\\
&+\sin\theta\hat{\tau}_1\left(M_0-i\cot\theta\boldsymbol{M}\cdot\boldsymbol{\sigma} \right).
\end{split}
\end{equation}
The advantage of using this parameterization is that $\theta$ and $M_0$ are real scalars and $\boldsymbol{M}=(M_1,M_2,M_3)$ is a real vector in the Matsubara representation. The normalization condition $\check{g}_0^2=1$ adds the constraint
\begin{equation}\label{eq:Constrain0}
M_0^2-\vert\boldsymbol{M}\vert^2=1.
\end{equation}

\begin{widetext}
With the parameterization in Eq.~\eqref{eq:ParamOfg0}, we get a set of differential equations from Eq.~\eqref{eq:RotUsaEqu}
\begin{equation}\label{eq:DiffEqnSet1}
D\partial_x^2\theta+2M_0\left(\Delta\cos\theta-\omega_n\sin\theta \right)-2hM_3\cos\theta
-\frac{1}{4\tau_{sf}}\left(2M_0^2+1\right)\sin\left(2\theta \right)=0,
\end{equation}
\begin{equation}\label{eq:DiffEqnSet0}
\begin{split}
D\left(\boldsymbol{M}\partial_x^2M_0-M_0\partial_x^2\boldsymbol{M} \right)+&2DM_0\alpha'(x)
\begin{pmatrix}
-\partial_xM_3\\
0\\
\partial_xM_1
\end{pmatrix}
+DM_0\alpha'(x)^2
\begin{pmatrix}
M_1\\
0\\
M_3
\end{pmatrix}
+2\boldsymbol{M}\left(\Delta\sin\theta+\omega_n\cos\theta \right)\\
&-2hM_0\sin\theta
\begin{pmatrix}
0\\
0\\
1
\end{pmatrix}
+\left[\frac{1}{\tau_{so}}+\frac{1}{2\tau_{sf}}\cos\left(2\theta\right)\right]M_0\boldsymbol{M}=0.
\end{split}
\end{equation}
We can see directly from the vector differential equation \eqref{eq:DiffEqnSet0} that the parameter $M_2$ does not have a source term from the domain wall structure, and therefore $M_2=0$ everywhere. Furthermore, one cannot directly solve these equations without the help of the constraint in Eq.~\eqref{eq:Constrain0}. Therefore, it is more convenient to use this constraint to transform Eq.~\eqref{eq:DiffEqnSet0} to another set of differential equations for each of the component of $M_i$ as in Ref.~\onlinecite{PhysRevB.94.104518}. 

Taking the second derivative of the constraint in Eq.~\eqref{eq:Constrain0}, we obtain
\begin{equation}\label{eq:SecDevCons}
\left(\partial_xM_0 \right)^2-\left(\partial_xM_1 \right)^2-\left(\partial_xM_3 \right)^2
+M_0\partial_x^2M_0-M_1\partial_x^2M_1-M_3\partial_x^2M_3=0.
\end{equation}
Substituting Eq.~\eqref{eq:SecDevCons} to each component of Eq.~\eqref{eq:DiffEqnSet0}, yields
\begin{equation}
\begin{split}
\label{eq:DiffEqnSet2}
D\partial_x^2&M_j-2\alpha'(x)DM_j(M_3\partial_xM_1-M_1\partial_xM_3)+DM_j\sum_{k=0,1,3}(-1)^k\left(\partial_xM_k \right)^2-\alpha'(x)^2DM_j(M_1^2+M_3^2)\\
&-2M_jM_0\left(\right.\omega_n\cos\theta+\Delta\sin\theta\left.\right)+2hM_jM_3\sin\theta-\left[\frac{1}{\tau_{so}}+\frac{1}{2\tau_{sf}}\cos\left(2\theta\right)\right]M_j\left(M_0^2-\delta_{j0}\right)=S_j,
\end{split}
\end{equation}
where $j=0,1,3$, and 
$$
S_j=\left\lbrace\begin{matrix}
-2\left(\omega_n\cos\theta+\Delta\sin\theta \right) && j=0\\
-2\alpha'(x)D\partial_xM_3+\alpha'(x)^2DM_1 && j=1\\
2\alpha'(x)D\partial_xM_1+\alpha'(x)^2DM_3-2h\sin\theta && j=3
\end{matrix}\right..
$$

The differential equations in (\ref{eq:DiffEqnSet1}) and (\ref{eq:DiffEqnSet2}) have to be supplemented by the boundary conditions in Eq.~\eqref{eq:BoundCond} and the solutions at the regions far from the domain wall. With the parameterization in Eq.~\eqref{eq:ParamOfg0}, Eq.~\eqref{eq:BoundCond} becomes
\begin{align}
\partial_x\theta\bigg\vert_{x_b^{\pm}}=\partial_x\theta\bigg\vert_{x_b^{\mp}}
\begin{split}
\left(M_i\partial_xM_0-M_0\partial_xM_i \right)\bigg\vert_{x_b^{\pm}}-\alpha'(x)M_0M_{i'}\bigg\vert_{x_b^{\pm}}
=\left(M_i\partial_xM_0-M_0\partial_xM_i \right)\bigg\vert_{x_b^{\mp}},\ i=1,3,
\end{split}
\end{align}
where $M_{i'}=M_3$ for $i=1$, and $M_{i'}=-M_1$ for $i=3$. 
\end{widetext}

\subsection{Solutions of the Usadel equation}

\begin{figure}[t]
\includegraphics[width=1.0\linewidth]{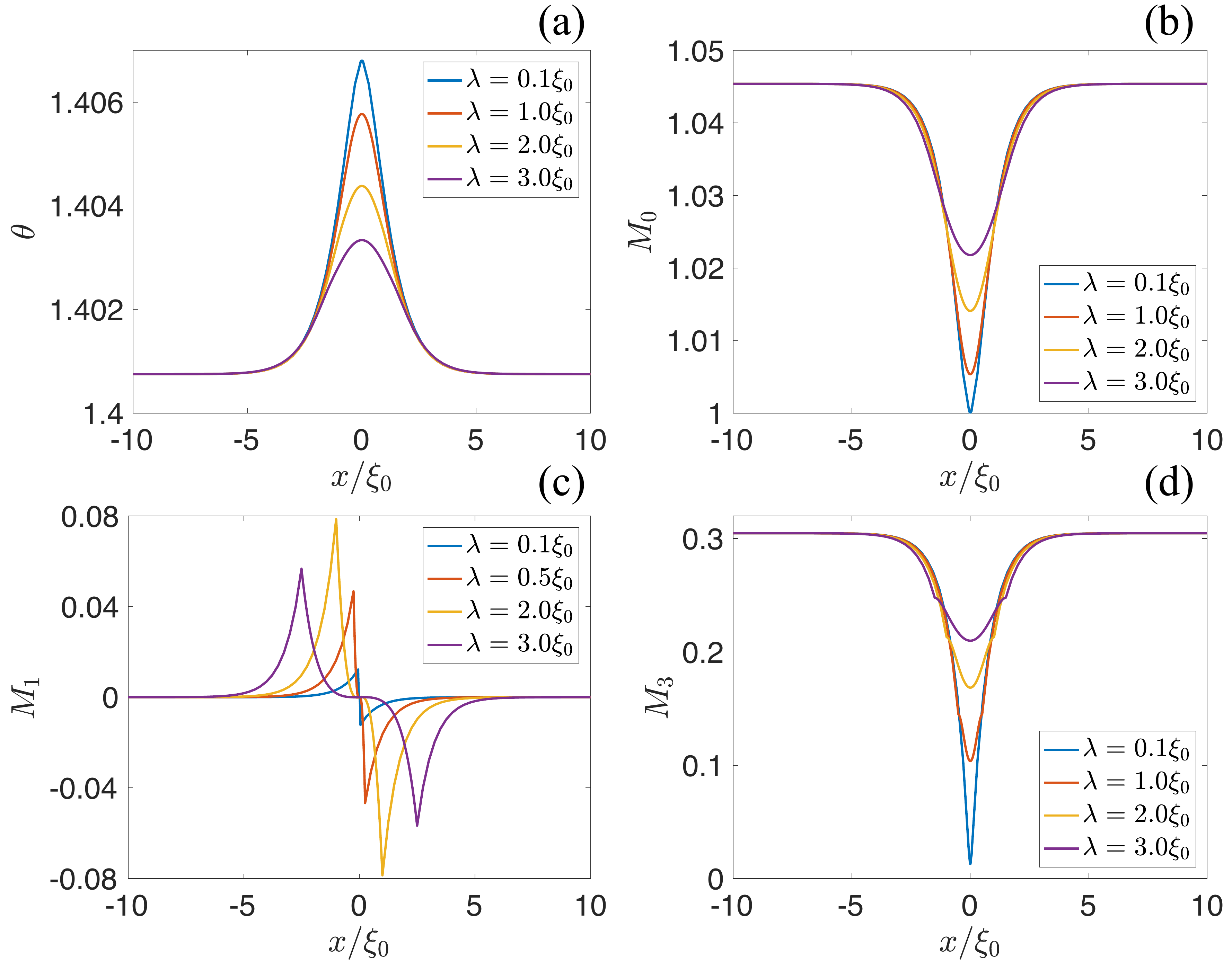}% Here is how to import EPS art
\caption{\label{fig:solutions} Solutions of the differential equations in Eqs.~\eqref{eq:DiffEqnSet1} and \eqref{eq:DiffEqnSet2} for different sizes of the domain wall. Here $\boldsymbol{h}=h(\cos\alpha(x),0,\sin\alpha(x))$, $h=0.3\Delta_0$, $\omega_n=\pi T$, $T=0.05\Delta_0$, $\tau_{so/sf}^{-1}=0$, $\Delta_0$ is the superconducting pair potential at zero temperature and exchange field, and $\xi_0=\sqrt{D/\Delta_0}$ is the superconducting coherence length .}
\end{figure}

Without the exchange field $h=0$, the solutions of the differential equations (\ref{eq:DiffEqnSet1}) and (\ref{eq:DiffEqnSet2}) give the regular results of the $\theta$ parameterization with $M_0=1$ and $M_i=0$. In the case of a homogeneous exchange field $\boldsymbol{h}=h\hat{z}$, the values of $\theta$ and $M_0$ are changed, and $M_3\neq0$, due to the odd-frequency spin triplet superconductivity (with zero spin projection on the spin $z$ axis) is induced. In the absence of spin relaxation terms
\begin{align*}
\tan&\theta=\frac{\sqrt{4\omega_n^2\Delta^2+\left(h^2+\omega_n^2-\Delta^2\right)^2}-h^2-\omega_n^2+\Delta^2}{2\omega_n\Delta}\\
&M_0=\frac{\omega_n+\Delta\tan\theta}{\sqrt{\omega_n^2+\left( \Delta^2-h^2\right)\tan^2\theta+2\omega_n\Delta\tan\theta}}\\ 
&M_3=\frac{h\tan\theta}{\sqrt{\omega_n^2+\left( \Delta^2-h^2\right)\tan^2\theta+2\omega_n\Delta\tan\theta}}\\
&M_1=0.
\end{align*}
These results can be used to describe the solution of the differential equations far from the domain wall structure. For the inhomogeneous exchange field with a domain wall structure, however, the differential equations (\ref{eq:DiffEqnSet1}) and (\ref{eq:DiffEqnSet2}) cannot be solved analytically, but the numerical solutions are plotted in Fig.~\ref{fig:solutions}.

We can see that the domain wall structure brings certain changes to the homogeneous solutions. For $\theta$, $M_0$, and $M_3$, the effect of the exchange field around the domain wall is reduced. This introduces a Gaussian function like structure for the solutions of these parameters as a function of position. At the center of the domain wall, the effect of the exchange field is reduced the most and the values of these parameters reach their homogeneous solutions gradually away from the domain wall. For a smaller domain wall, the reduction of the effect of the exchange field is more obvious, until for $\lambda\rightarrow0$, the effect of the exchange field is completely lifted at the center of the domain wall. This is due to the existence of the spin-orbit relaxation like term in Eq.~\eqref{eq:SpiOrbRelLikTer} in the rotated Usadel equation with a relaxation rate $\alpha'(x)^2D/2$ and $\alpha'(x)\propto \lambda^{-1}$. 

The domain wall structure also introduces a nonzero solution of $M_1$ around the domain wall, due to the odd-frequency spin triplet superconductivity (with nonzero spin projection on the spin $z$ axis) induced by the inhomogeneous exchange field. The maximum of $M_1$ appears at the boundary of the domain wall and gradually goes to zero away from it. It also changes sign at the two sides of the domain wall center. The amplitude of $M_1$ first increases and then decreases with increasing $\lambda$, the maximum taking place at $\lambda\approx2\xi_0$. 

\begin{figure}[t]
\includegraphics[width=1.0\linewidth]{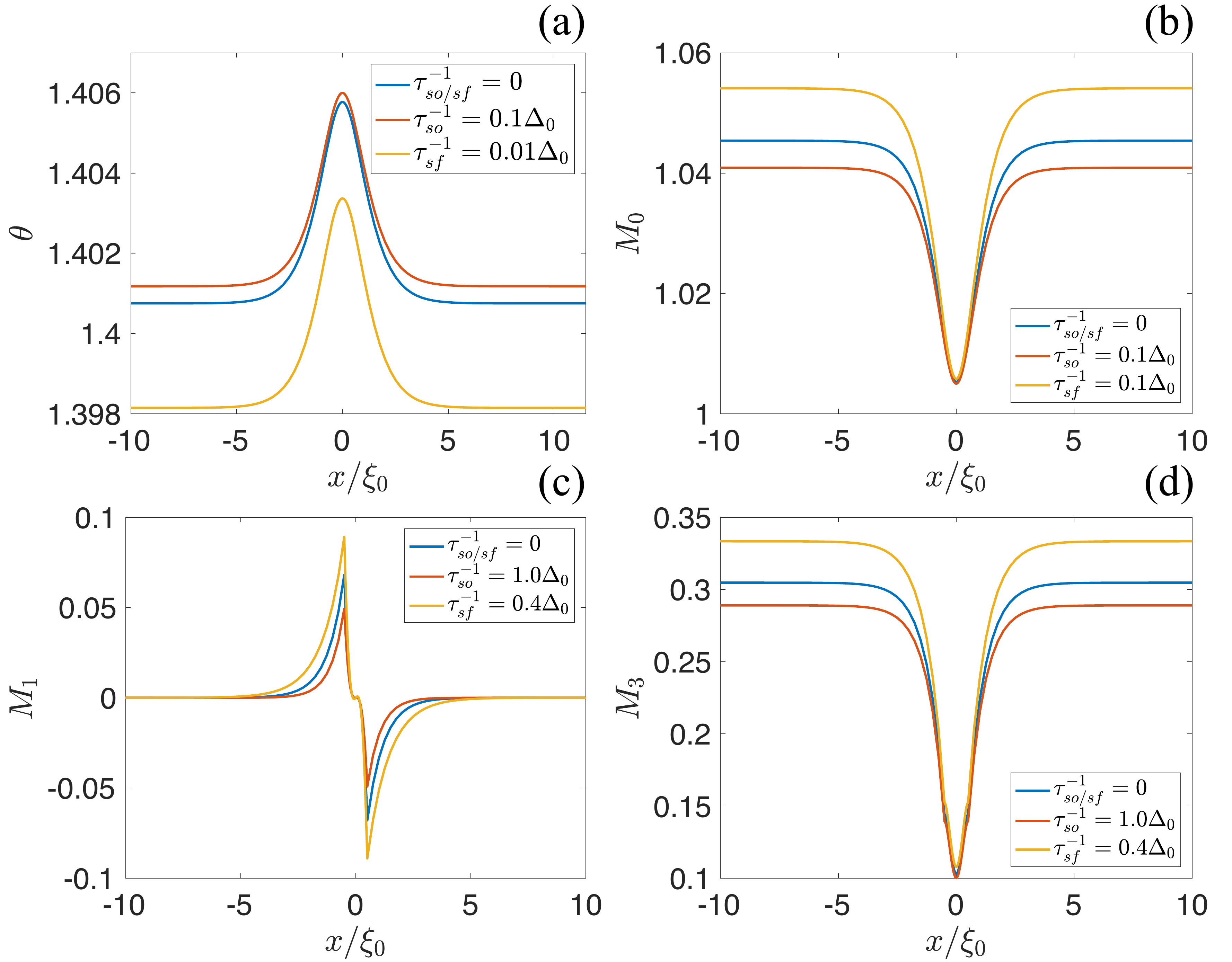}% Here is how to import EPS art
\caption{\label{fig:solutions2} Effect of spin relaxation on the solutions of the Usadel equation. Here $\boldsymbol{h}=h(\cos\alpha(x),0,\sin\alpha(x))$, $h=0.3\Delta_0$, $\omega_n=\pi T$, $T=0.05\Delta_0$, and $\lambda=1.0\xi_0$. }
\end{figure}

Spin relaxation also brings many changes to the solutions of the differential equations. For the regions far from the domain wall, the consequence of spin relaxation is the same as with the case of homogeneous exchange field. The spin-orbit relaxation reduces the effect of the exchange field, therefore, the solutions in these regions approach the ones in the domain wall center. The spin-flip relaxation reduces the superconducting pair potential $\Delta$, and therefore $\theta$ in these regions becomes smaller, and $M_{0/3}$ becomes larger with spin-flip relaxation. These can be seen from Fig.~\ref{fig:solutions2}(a,b,d).  

For the regions around the domain wall, the spin relaxation brings minor changes. This is due to the fact that at the center of the domain wall, the effect of the exchange field is already reduced, and spin relaxation affects superconductivity similarly to the case without exchange field. This can be seen in Fig.~\ref{fig:solutions2}(b,d). These behaviors can also be revealed in the physical quantities as discussed below. 

With these solutions, and using Eq.~\eqref{eq:FullGreensFunction}, the unrotated Green's function is
\begin{equation}\label{eq:ParamOfg}
\begin{split}
\check{g}&=\cos\theta\hat{\tau}_3\left[M_0+i\tan\theta\cos\alpha(x)\boldsymbol{M}\cdot\boldsymbol{\sigma}\right.\\
&\left.+i\tan\theta\sin\alpha(x)\left(M_3\hat{\sigma}_1-M_1\hat{\sigma}_3\right)\right]\\
&+\sin\theta\hat{\tau}_1\left[M_0-i\cot\theta\cos\alpha(x)\boldsymbol{M}\cdot\boldsymbol{\sigma} \right.\\
&\left.-i\cot\theta\sin\alpha(x)\left(M_3\hat{\sigma}_1-M_1\hat{\sigma}_3\right) \right].
\end{split}
\end{equation}
In the rest of the paper, we use this Green's function to calculate various physical quantities. 

\subsection{Self-consistent calculations}
The superconducting pair potential in Eq.~\eqref{eq:RotUsaEqu} has to be determined self-consistently. In the matsubara representation, we have
\begin{equation}
\label{eq:selfconsg}
\Delta=\frac{1}{2}\pi T\gamma \sum_{\omega_n>0}^{\omega_D}\text{Tr}\left(\hat{\tau}_1\check{g} \right)=2\pi T\gamma\sum_{\omega_n>0}^{\omega_D}M_0\sin\theta,
\end{equation}
where $\check{g}$ is given by Eq.~\eqref{eq:ParamOfg}, $\gamma$ is the coupling constant, and $\omega_D$ is the BCS cutoff energy. The latter gives a temperature dependent cutoff $N_D=\omega_D/(2 \pi T)$ to the sum over $n$. Considering the relations
\begin{equation}
\label{eq:weakcouplingresult}
    \begin{split}
2\pi T\sum_{n=0}^{N_D(T)}\frac{1}{\omega_n}&=2\pi T\left( \sum_{n=0}^{N_D(T_c)}+\sum_{N_D(T_c)}^{N_D(T)}\right)\frac{1}{\omega_n}
\\
&=\frac{1}{\gamma}+\log\left(\frac{T_c}{T}\right)
\end{split}
\end{equation}
we can rewrite the self-consistency equation as
\begin{equation}\label{eq:SelfConEqn}
\Delta\log\left(\frac{T}{T_c} \right)=2\pi T\sum_{\omega_n>0}\left(M_0\sin\theta-\frac{\Delta}{\omega_n} \right).
\end{equation}
We can see that $\Delta$ does not explicitly depend on the rotation angle $\alpha(x)$. Its position dependence comes from the parameters $M_0$ and $\theta$, whose solutions depend on $\alpha(x)$. 
This is because we only consider spin-singlet pairing in the self-consistent calculations. 

The self-consistent pair potential in Eq.~\eqref{eq:SelfConEqn} is position dependent, $\Delta=\Delta(x)$. At the regions far from the domain wall, $\Delta(x=\pm\infty)$ is the same as in the case of homogeneous exchange field. A homogeneous exchange field brings many interesting effects on $\Delta$. It suppresses $\Delta$ at a finite temperature and for $T\rightarrow0$, $\Delta\rightarrow0$ for a field $h>h_c$. Here $h_c=\Delta_0/\sqrt{2}$ when $T=0$, called the Chandrasekhar-Clogston limit\cite{chandrasekhar1962note,clogston1962upper}. The superconducting pair potential is reduced by spin relaxation, and $h_c$ is enhanced by the spin-orbit relaxation and suppressed by the spin-flip relaxation.\cite{RevModPhys.90.041001,2017arXiv170608245B} 

In the case of an inhomogeneous exchange field, above properties are similar. However, in the domain wall region, the weak effect of the spin relaxation (Fig.~\ref{fig:solutions2}) brings less changes to $\Delta$, compared to the homogeneous case. 

\section{Equilibrium Spin Current Density\label{sec:EquilibriumSpinCurrentDensity}}
%\subsection{}
%\subsubsection{}
One important consequence of the inhomogeneous exchange field is the equilibrium spin current density. Due to the inhomogenetity of the magnetization, the spin of the quasiparticles rotates following the local magnetization, which creates a flow of spin\cite{ralph2008spin}. 

In the quasiclassical theory, the spin current can be calculated from
$$
j_{k,i}=\frac{\sigma_N}{2e^2}\pi Ti\sum_{\omega_n>0}\text{Tr}\left[\hat{\tau}_0\hat{\sigma}_i\left(\check{g}\nabla_k\check{g}\right) \right],
$$
where $\sigma_N=2e^2N_0D$ is the normal state conductivity and $N_0$ is the density of states at the Fermi level. The spin current density $j_{k,i}$ is a tensor, the index $k$ represents the transport direction of spin, and $i$ represents the spin component. With the parameterization in Eq.~\eqref{eq:ParamOfg}, we get the three components of the spin current density in the $x$ direction,
\begin{align}\label{eq:SpinCurrDen}
\begin{split}
j_{x,2}=\frac{2\sigma_N}{e^2}\pi T\sum_{\omega_n>0}&\left[\alpha'(x)(M_1^2+M_3^2)\right.\\
&\left.+M_3\partial_xM_1-M_1\partial_xM_3 \right]
\end{split}
\end{align}
$$
j_{x,1/3}=0,
$$
where $\alpha(x)$ is the rotation angle in Eq.~\eqref{eq:RotationAngle}. Since the rotation of the magnetization happens in the $xz$ plane, the $x$ and $z$ spin components of the spin current density are absent, $j_{x,1/3}=0$. The only nonzero spin component of the spin current density is in the $y$ spin direction. 

\begin{figure}[t]
\includegraphics[width=1.0\linewidth]{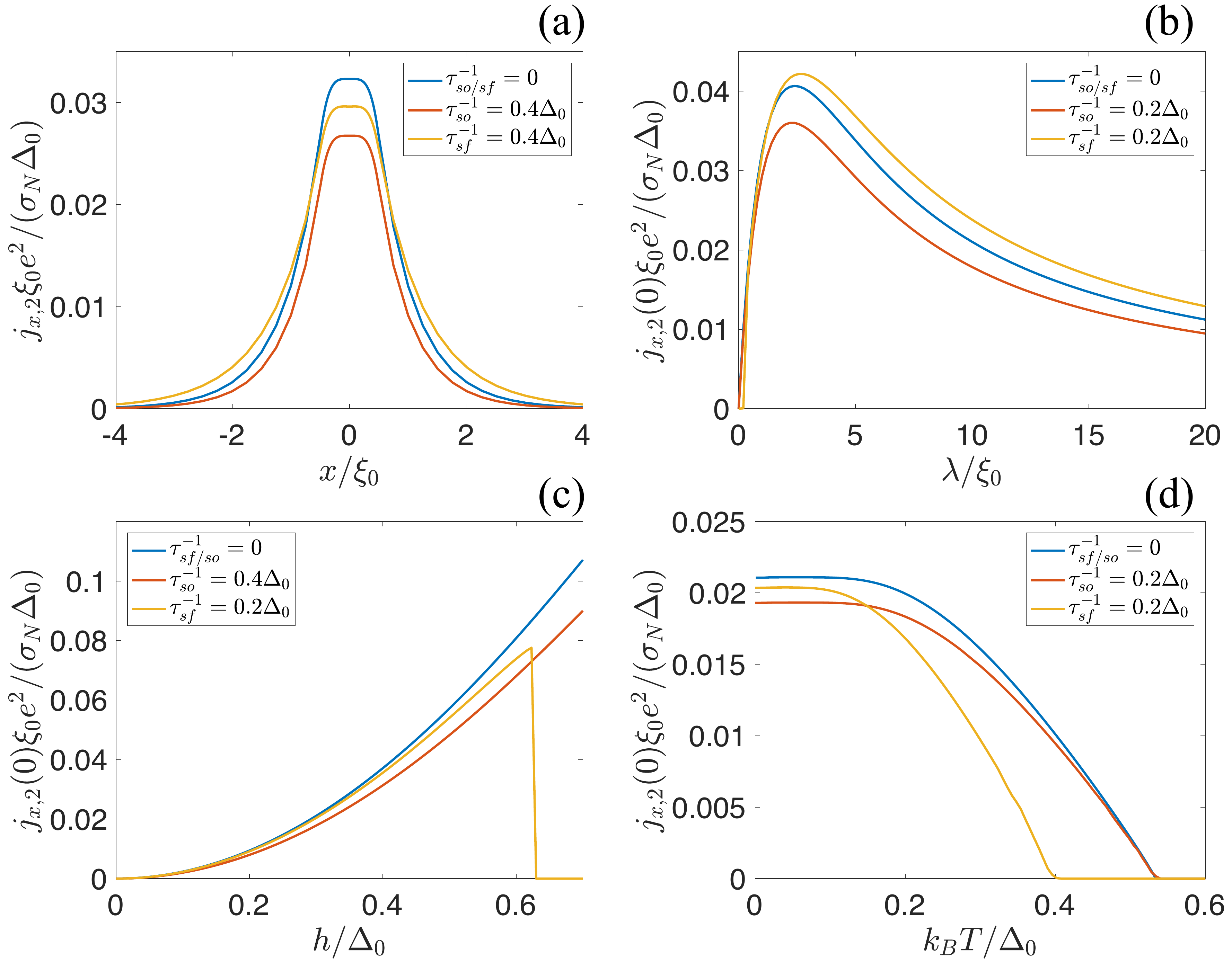}% Here is how to import EPS art
\caption{\label{fig:SpinCurrent} Dependence of spin current density on (a) position, and the maximum spin current density $j_{x,2}(0)$ on (b) domain wall size, (c) exchange field, and (d) temperature. Here $\lambda=0.5\xi_0$, $T=0.05\Delta_0$ and $h=0.3\Delta_0$ unless specified otherwise.}
\end{figure}

The spatial dependence of $j_{x,2}$ is shown in Fig.~\ref{fig:SpinCurrent}(a) for a domain wall size $\lambda=0.5\xi_0$. We can see that a nonzero spin current density is created around the domain wall structure. Inside the domain wall, the maximum spin current density $j_{x,2}(0)$ is constant, and smoothly goes to zero outside the domain wall. Both spin-orbit and spin-flip relaxations reduce the spin current density.

The constant spin current density $j_{x,2}(0)$ in the domain wall region is determined from the adiabatic solution of the parameter $M_3$ and the domain wall size $\lambda$. From Fig.~\ref{fig:solutions}(d) we know that $M_3$ describes the effect of the exchange field on superconductivity, and at the center of the domain wall $M_3\rightarrow0$ for a small domain wall size. Therefore, $j_{x,2}(0)\rightarrow0$ for $\lambda\rightarrow0$ and becomes large for a stronger exchange field. These are shown in Fig.~\ref{fig:SpinCurrent}(b,c). We can also see that for a small domain wall, both spin-orbit and spin-flip relaxation reduce $j_{x,2}(0)$. However, for large domain walls, spin-flip relaxation increases $j_{x,2}(0)$ for a fixed domain wall size, but it kills $j_{x,2}(0)$ at the Chandrasekhar-Clogston limit, where the superconductivity is suppressed\cite{RevModPhys.90.041001,2017arXiv170608245B}.

The temperature dependence of $j_{x,2}(0)$ is shown in Fig.~\ref{fig:SpinCurrent}(d). It resembles the temperature dependence of the self-consistent pair potential $\Delta$. This means $j_{x,2}(0)\sim\Delta$ for a fixed domain wall size and exchange field $h$. The effect of spin relaxation also resembles  the temperature dependence of $\Delta$.

From the spatial dependence of $j_{x,2}$ we can see that, strictly speaking, the spin current density is not conserved. The position dependence is related with the appearance of a spin-transfer torque\cite{ralph2008spin}. It is exerted on the spins in order to reorient spin flow to follow the direction of local magnetization, namely, it represents the rotation of spins through the domain wall structure. Then in the continuity equation for spin density $\boldsymbol{n}_s$, we have \cite{sun2005definition}
$$
\frac{d\boldsymbol{n}_s}{dt}+\nabla\cdot j_{k,i}=\mathcal{T},
$$
\\
where $\mathcal{T}$ is the spin-transfer torque and it is nonzero in the case of an inhomogeneous exchange field and spin-relaxation.

In the static case $\nabla\cdot j_{k,i}=\mathcal{T}$. With the parameterization in Eq.~\eqref{eq:ParamOfg},
\begin{equation}\label{eq:DefinitionOfSTT}
\mathcal{T}=\partial_x j_{x,2}=\frac{4\sigma_N}{e^2D}\pi Th\sum_{\omega_n>0}\sin\theta M_1.
\end{equation}
It is nothing but the Matsubara sum of $M_1$ but constrained by superconductivity ($\sin\theta$). In the normal state $\theta\rightarrow0$, then the torque is zero in equilibrium. 

The expression of $\mathcal{T}$ in Eq.~\eqref{eq:DefinitionOfSTT} also helps us to understand the properties of the spin current density. The spatial dependence follows Fig.~\ref{fig:solutions}(c), but the amplitude is constrained by $\sin\theta\sim\Delta$. That is why the temperature dependence of the spin current density is similar to that of the self-consistent pair potential (Fig.~\ref{fig:SpinCurrent}(d)). The dependence on $h$ explains the monotonous dependence on $h$ of the spin current density (Fig.~\ref{fig:SpinCurrent}(c)). 

\begin{widetext}
The equilibrium spin-transfer torque $\mathcal{T}$ is related to the superconducting free energy by
\begin{equation}
\label{eq:RelationOfTorqueAndEnergy}
\mathcal{T}=\frac{1}{V}\boldsymbol{h}\times\frac{\delta F_{sn}}{\delta \boldsymbol{h}},
\end{equation}
where $V$ is the volume of the superconductor, $\boldsymbol{h}$ is the exchange field induced in the superconductor, and $F_{sn}$ is given by
\begin{equation}\label{eq:SupFreEne}
F_{sn}=W\int_{-\infty}^{\infty}f_{sn}dx,
\end{equation}
where $W$ is the cross sectional area of the superconductor and the superconducting free energy density\cite{Eilenberger1968,PhysRevLett.25.507,Altland1998,PhysRevB.64.014512,PhysRevB.96.245311} compared to its normal state $f_{sn}$ is given by
(see Appendix~\ref{app:fsn})
\begin{equation}
\label{eq:fsndef}
\begin{split}
f_{sn}=f_s-f_n&=\pi TN_0\sum_{\omega_n>0}\text{Tr}\left\{ \left(\omega_n+i\boldsymbol{h}\cdot\boldsymbol{\sigma} \right)\left(\check{1}-\hat{\tau}_3\check{g}\right) \right.-\frac{1}{2}\left(\Delta\hat{\tau}_++\Delta^*\hat{\tau}_- \right)\check{g}+\frac{D}{4}\left(\nabla\check{g} \right)^2\\
&+\frac{1}{16\tau_{so}}\left[3-\left( \boldsymbol{\sigma}\check{g}\right)\cdot\left( \boldsymbol{\sigma}\check{g}\right) \right]+\frac{1}{16\tau_{sf}}\left[3-\left( \boldsymbol{\sigma}\hat{\tau}_3\check{g}\right)\cdot\left( \boldsymbol{\sigma}\hat{\tau}_3\check{g}\right) \right] \}.
\end{split}
\end{equation}
In the absence of spin relaxation and exchange field, this agrees with the result in Ref.~\onlinecite{PhysRevB.96.245311}. Here $f_{sn}=f_{sn}[\boldsymbol{h}]$, and $\Delta$ and $\check{g}$ are the self-consistent values of the order parameter and the Green function. 
With the parameterization in Eq.~\eqref{eq:ParamOfg}, it becomes
\begin{align}\label{eq:SupFreEneDen}
\begin{split}
f_{sn}&=\pi TN_0\sum_{\omega_n>0}\Big\{ 4\omega_n-2M_0\left(2\omega_n\cos\theta+\Delta\sin\theta \right)+4hM_3\sin\theta-D\left(M_1^2+M_3^2\right)\alpha'(x)^2\\
&+D\left[\left(\partial_xM_0\right)^2-\left(\partial_xM_1\right)^2-\left(\partial_xM_3\right)^2+\left(\partial_x\theta\right)^2\right]-2D\left(M_3\partial_xM_1-M_1\partial_xM_3\right)\alpha'(x)\\
&+\frac{1}{4}\left[3\left(\tau_{so}^{-1}+\tau_{sf}^{-1}\right)-3\left(\tau_{so}^{-1}+\tau_{sf}^{-1}\cos{2\theta} \right)M_0^2 -\left(\tau_{so}^{-1}-\tau_{sf}^{-1}\cos{2\theta} \right)\left(M_1^2+M_3^2 \right) \right] \Big\}.
\end{split}
\end{align}
\end{widetext}

From the relation in Eqs.~(\ref{eq:SpinCurrDen},\ref{eq:RelationOfTorqueAndEnergy}), we obtain
\begin{equation}\label{eq:FEdensityWithSpinCurr}
f_{sn}=f_{sn}^0-\frac{1}{2}\alpha'(x)j_{x,2}+\frac{1}{4}\alpha'(x)^2\frac{\delta j_{x,2}}{\delta\alpha'(x)},
\end{equation}
where $f_{sn}^0$ is the free-energy density in Eq.~\eqref{eq:SupFreEneDen} with terms that do not directly depend on $\alpha'(x)$, and $f_{sn}-f_{sn}^0$ is nonzero only in the domain wall region. We can see that the spin current density contributes to the energetics of the system, which in turn influences the formation of the domain wall. 

In nonequilibrium spin transport, spin-transfer torque leads to the domain wall motion and influences the orientation of the magnetization\cite{ralph2008spin}. In our model, the equilibrium spin-transfer torque in Eq.~\eqref{eq:DefinitionOfSTT} does not make the domain wall move, but it contributes to the superconducting free energy via the spin current density (as in  Eq.~\eqref{eq:FEdensityWithSpinCurr}), which in turn affects the domain wall size, as we discuss in the next section.

\section{Effect of superconductivity on the domain wall size\label{sec:Effectofsuperconductivityonthedomainwallsize}}

In the absence of the external magnetic field, the domain wall size in a ferromagnet is dictated by the competition between exchange and anisotropy energies\cite{landau2013electrodynamics}. Exchange energy tries to maintain the direction of the magnetization, while the anisotropy energy tends to align the magnetization to a particular direction.

The exchange energy density can be expressed by
$$
U_{ex}=Q\sum_{i=1}^{3}\left(\frac{\partial m_i}{\partial x} \right)^2,
$$
where $m_i$ is the component of the magnetization unit vector,  and $Q$ is the exchange stiffness constant. Here $\boldsymbol{m}=(\cos\alpha,0,\sin\alpha)$, where $\alpha$ is the rotation angle in Eq.~\eqref{eq:RotationAngle}. Substituting Eq.~\eqref{eq:RotationAngle} to $U_{ex}$ we get
\begin{equation}\label{eq:ExEnDen}
U_{ex}=\frac{Q\pi^2}{\lambda^2}\Theta\left(x+\frac{\lambda}{2}\right)\Theta\left(\frac{\lambda}{2}-x\right).
\end{equation}

The anisotropy energy density depends on the crystal structure of the system. Most of the ferromagnetic insulators have face centered cubic crystal structure\cite{doi:10.1063/1.4866265}. The anisotropy energy density in this case is given by
$$
U_{\textrm{aniso}}=K_{c1}\left(m_1^2m_2^2+m_1^2m_3^2+m_2^2m_3^2\right)+K_{c2}m_1^2m_2^2m_3^2,
$$
where $K_{c1},K_{c2}<0$ for many ferromagnetic insulators, which makes the magnetization lie in one of the easy planes. In the case of thin films, the symmetry is broken in the direction perpendicular to the film plane and this energy density can be expressed by a uniaxial crystal structure  as\cite{PhysRevLett.115.087201} 
\begin{equation}\label{eq:AnEnDen}
U_{\textrm{aniso}}=\left(K_1+\frac{K_s}{t} \right)\sin^2\alpha+K_2\sin^4\alpha,
\end{equation}
where $K_1=K_{c1}$, $K_2=-7K_{c1}/8+K_{c2}/8$, and $K_s$ is the surface anisotropy constant representing the rotation of the easy plane towards an easy axis magnetization with film thickness $t$. 

Together with Eqs.~(\ref{eq:ExEnDen},\ref{eq:AnEnDen}), the domain wall energy is given by
\begin{equation}\label{eq:FreeEneInFerr}
F_1=\int_{-\infty}^{\infty}\left(U_{ex}+U_{\textrm{aniso}} \right)dx=\frac{Q\pi^2}{\lambda}+\frac{1}{2}K_{\textrm{eff}}\lambda,
\end{equation}
where $K_{\textrm{eff}}=K_1+K_s/t+3K_2/4$. Minimization of this energy with respect to $\lambda$ gives the domain wall size of the inhomogeneous exchange field in the ferromagnet
\begin{equation}\label{eq:DomWalSizInFerr}
\lambda_0=\sqrt{\frac{2Q}{K_{\textrm{eff}}}}\pi.
\end{equation}
Then the minimized domain wall energy is 
\begin{equation}\label{eq:MinFreeEne}
F_1^{\textrm{min}}=\sqrt{2K_{\textrm{eff}}Q}\pi.  
\end{equation}

In many studies, the domain wall structure is represented by the hyperbolic functions with~\cite{landau2013electrodynamics}
$$
\alpha(x)=\cos^{-1}\left[-\tanh\left(\frac{x-x_0}{\lambda'} \right) \right],
$$
where $\lambda'=\sqrt{Q/K_{\textrm{eff}}'}$ and $K_{\textrm{eff}}'=K_1+K_s/t+2K_2/3$. The minimized energy is then given by $4\sqrt{K_{\textrm{eff}}'Q}$. Since $K'_{\textrm{eff}}\approx K_{\text{eff}}$, this is very close to $F_1^{\textrm{min}}$ in Eq.~\eqref{eq:MinFreeEne}. In other words, the domain wall structure in Eq.~\eqref{eq:RotationAngle} gives the minimum energy which is only slightly larger with that of more complicated domain wall structures. 

\begin{figure}[t]
\includegraphics[width=1.0\linewidth]{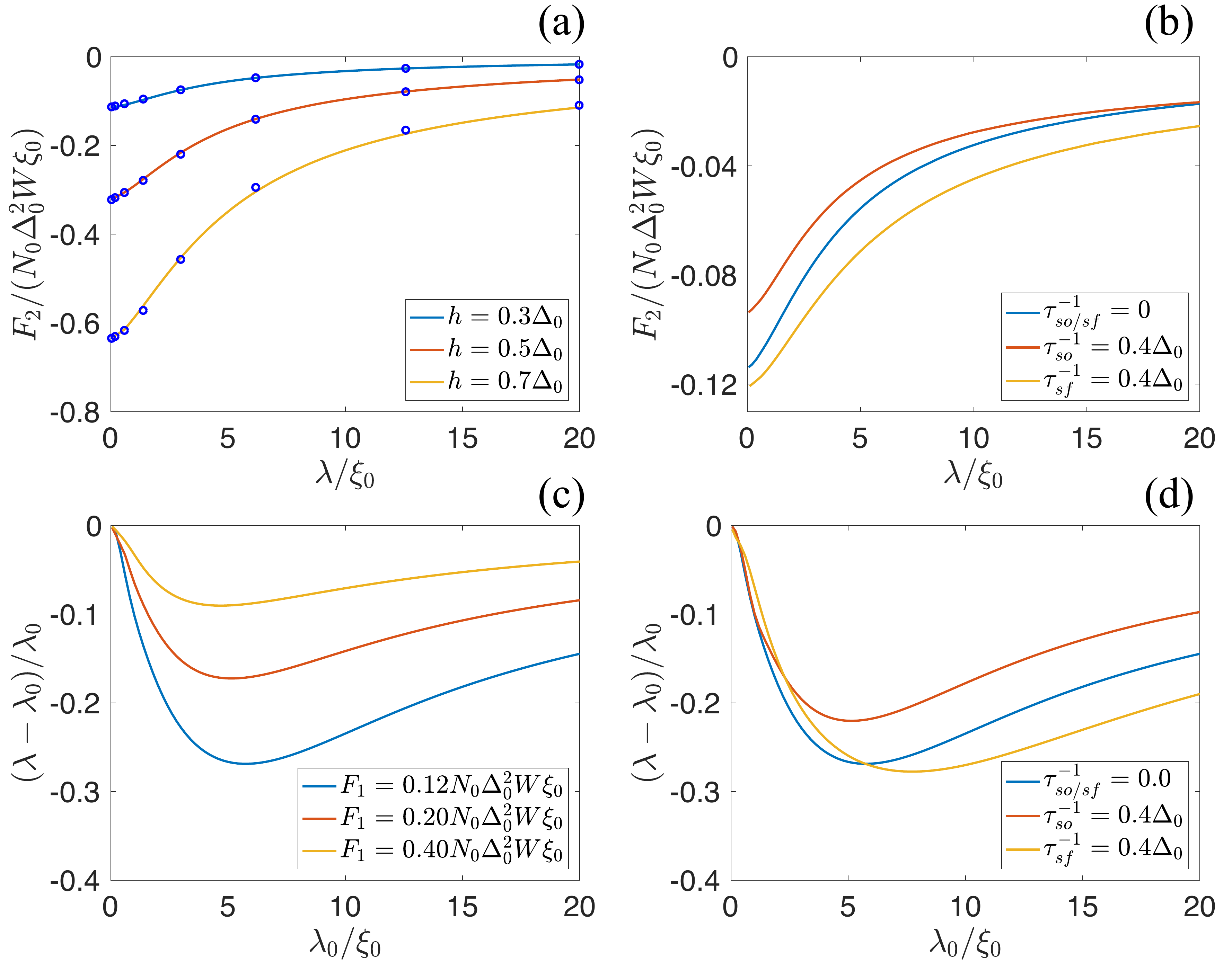}% Here is how to import EPS art
\caption{\label{fig:DomainWallSize} (a) Superconducting free energy contribution $F_2$ to the domain wall energy at $T=0.05\Delta_0$ for different exchange field strengths. The blue circles in (a) are the results of the fitted formula in Eq.~\eqref{eq:FreEneConSup}. (b) Effect of spin relaxation on $F_2$ for $h=0.3\Delta_0$ at $T=0.05\Delta_0$. (c) Effect of superconductivity on the domain wall size for $h=0.3\Delta_0$ with different domain wall energies $F_1$ in the ferromagnet. (d) The effect of spin relaxation on the same quantity as in (c) for $F_1=0.12N_0\Delta_0^2W\xi_0$.}
\end{figure}

The contribution of superconductivity to the domain wall energy is given by the difference of the free energy in the cases of inhomogeneous and homogeneous magnetization
$$
F_2=F_{sn}(\boldsymbol{h})-F_{sn}(\boldsymbol{h}=h\hat{z}),
$$
where $F_{sn}$ is given by Eq.~\eqref{eq:SupFreEne}. 
This energy cannot be expressed analytically, but the numerical result is plotted in Fig.~\ref{fig:DomainWallSize}(a). We can see that $F_2$ is negative. This is due to the fact that the existence of the domain wall structure enhances the superconducting condensation energy near the domain wall. This contribution is stronger for a smaller domain wall, since smaller domain wall reduces the effect of the exchange field more in the domain wall region. The effect of spin relaxation is plotted in Fig.~\ref{fig:DomainWallSize}(b). We can see that spin-orbit relaxation makes $|F_2|$ smaller. This is because spin-orbit relaxation reduces the effect of the exchange field, but it has less effect in the domain wall region. This makes the free energy difference $|F_{sn}(\boldsymbol{h})-F_{sn}(\boldsymbol{h}=h\hat{z})|$ smaller. Spin-flip relaxation on the other hand, makes $|F_2|$ larger.

From Fig.~\ref{fig:DomainWallSize}(a) we can approximate $F_2$ as
\begin{equation}\label{eq:FreEneConSup}
   F_2=-\frac{a}{\lambda+2\xi_0}+be^{-\lambda/\xi_0}, 
\end{equation}
where $a,b$ are functions of temperature and exchange field strength and can be determined numerically. For example, for $h=0.3\Delta_0$, $a=0.4N_0\Delta_0^2W\xi_0^2$ and $b=0.085N_0\Delta_{0}^2W\xi_0$. 

Minimizing the total energy $F_1+F_2$ with respect to $\lambda$, we get a compact analytical expression for two extreme limits\\
(i) $\lambda\ll\xi_0$
$$
\lambda=\sqrt{\frac{2Q}{K_{\textrm{eff}}+\left(\frac{a}{2\xi_0^2}-\frac{2b}{\xi_0}\right)}}\pi.
$$
(ii) $\lambda\gg\xi_0$
\begin{equation}\label{eq:condition}
\lambda=\sqrt{\frac{2\left(Q-a/\pi^2\right)}{K_{\textrm{eff}}}}\pi,\ \text{for}\ Q\pi^2>a
\end{equation}
$$
\lambda=\frac{4a\xi_0}{a-\pi^2Q},\ \text{for}\ Q\pi^2<a.
$$
In both limits superconductivity reduces the domain wall size. (i) For a small domain wall $\lambda\ll\xi_0$, superconductivity reduces the domain wall size by effectively increasing the anisotropy constant $K_{\textrm{eff}}$. (From the numerical results $a/(2\xi_0^2)>2b/\xi_0$ holds for all $h$.) (ii) For the case of a large domain wall, the situation is more complicated. For a ferromagnet with strong stiffness $Q\pi^2>a$, superconductivity reduces the domain wall size by effectively reducing the exchange stiffness constant $Q$. For a ferromagnet with weak stiffness $Q\pi^2<a$, which also refers to the case of $K_{\textrm{eff}}\ll1.0N_0\Delta_{0}^2W$, the domain wall size is also reduced. However, in this case superconductivity dominates the domain wall energy, and leads to a negative total energy $F_1+F_2$, which introduces a dense domain structure with a domain length comparable with the domain wall size.\cite{RevModPhys.77.935} Since we are considering a single domain structure, here we only consider domain walls with positive domain wall energy. 

In Fig.~\ref{fig:DomainWallSize}(c), we numerically minimize the domain wall energy $F=F_1+F_2$ with respect to the domain wall size for the case of  $h=0.3\Delta_0$ and $T=0.05\Delta_0$, and calculate the relative change of the domain wall size. To avoid a negative domain wall energy leading to a transition to a system with many domain walls, we set in each figure a constant $F_1\ge 0.12N_0\Delta_0^2 W \xi_0$ so that $F>0$ for each case considered. In other words, instead of varying $K_{\rm eff}$ and $Q$, we fix $F_1$ and vary $\lambda_0$ in the figures. The effect of superconductivity on the domain wall size is strongest for the lowest $F_1$. If we consider larger values of $F_1$, namely, a stronger ferromagnet, the effect of  superconductivity on the domain wall size is smaller.

The effect of spin relaxation on the domain wall size is shown in  Fig.~\ref{fig:DomainWallSize}(d). We see that for small domain walls, the spin relaxation brings very little effects on the domain wall size compared to the case without spin-relaxation. However, for larger domain walls, the two types of spin-relaxation mechanisms lead to different effects on the domain wall size. Spin-orbit relaxation makes the effect smaller since it reduces the effect of the exchange field and makes $|F_2|$ smaller. Spin-flip relaxation makes the effect stronger, since $|F_2|$ is larger.

As we can see, the effect of superconductivity on the domain wall size is pronounced for weak ferromagnets with large domain walls. Domain wall sizes in ferromagnetic insulators are rarely reported, but with Eq.~\eqref{eq:DomWalSizInFerr} we can evaluate a typical size. For materials with a face centered cubic structure, the exchange stiffness constant $Q$ is related to the exchange integral $J$ by $Q=4S^2J/a_0$, where $S$ is spin and $a_0$ is the lattice constant.~\cite{chikazumi2009physics} With the values of the parameters for $Q$ in Ref.~\onlinecite{PhysRevB.81.155213} and for $K_{\rm{eff}}$ in Ref.~\onlinecite{PhysRevLett.115.087201}, we get the domain wall size for a EuS thin film as $50\  \text{nm}$. Considering a typical coherence length of a conventional superconductor in the diffusive superconductors at zero temperature and exchange field between 15 nm and a few hundred nm, we can estimate in this case $\lambda_0\sim 0.1 \dots 5 \xi_0$. Therefore, although perhaps in most cases $\lambda_0 < \xi_0$ is the most relevant limit, nothing as such seems to exclude the possibility of the opposite limit as well. 

\section{Density of states\label{sec:Densityofstates}}
%\subsection{}
%\subsubsection{}
The inhomogeneous exchange field also makes the local density of states (DoS) peculiar. Since the domain wall structure reduces the effect of the exchange field, the local DoS in the domain wall region is different from the one for homogeneous magnetization. 

\begin{figure}[t]
\includegraphics[width=1.0\linewidth]{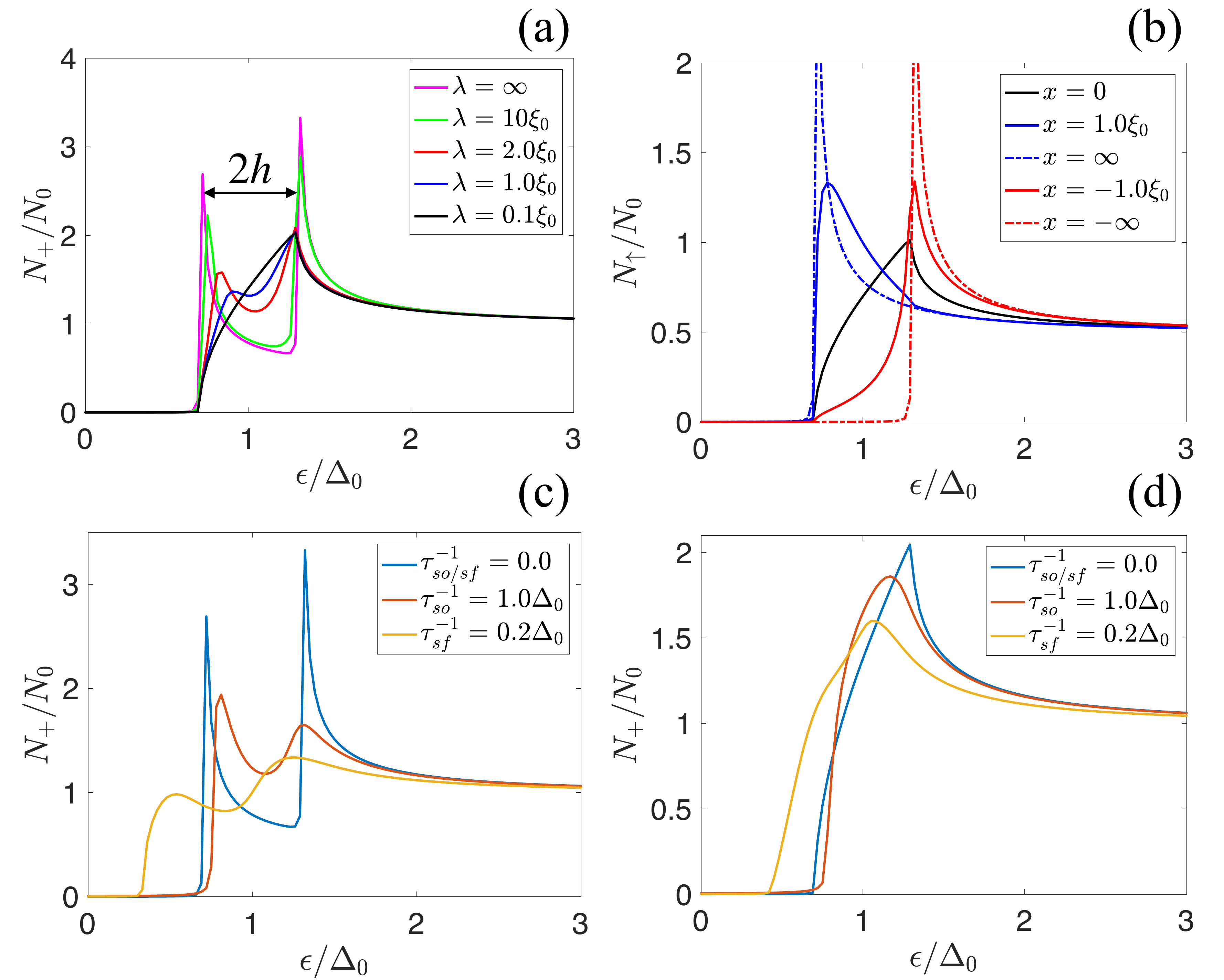}% Here is how to import EPS art
\caption{\label{fig:DensityOfStates} (a) Local DoS at the center of the domain wall for different domain wall sizes. (b) Local DoS for spin $\uparrow$ at different positions for a domain wall size $\lambda=0.1\xi_0$. (c) Local DoS in the presence of spin relaxation in the homogeneous exchange field (at $x=-\infty$) and (d) at the center of the domain wall (x=0) for a domain wall size $\lambda=0.1\xi_0$. The temperature and exchange field used in the calculations are $T=0.05\Delta_0$, $h=0.3\Delta_0$. }
\end{figure}

In the quasiclassical theory, the local DoS for each spin species is given by
$$
N_s=\frac{1}{8}N_0\text{Re}\left[\text{Tr}\left(\hat{\tau}_3\pm\hat{\sigma}_3\hat{\tau}_3\right)\check{g}\vert_{\omega_n=-i\epsilon^+} \right],
$$
where $\pm$ is for spin $\uparrow/\downarrow$. With the parameterization in Eq.~\eqref{eq:ParamOfg}, it becomes
\begin{equation}\label{eq:DoSForSpinS}
N_s=\frac{1}{2}N_0\text{Re}\left[\cos\theta M_0\pm\left(\cos\alpha M_3-\sin\alpha M_1 \right)i\sin\theta\right].
\end{equation}
In Fig.~\ref{fig:DensityOfStates}(a) we show the local DoS $N_+=N_{\uparrow}+N_{\downarrow}$ at the center of the domain wall for different domain wall sizes in the absence of spin relaxation. At the center of the domain wall $\alpha=\pi/2$ and $M_1=0$, then $N_\uparrow=N_{\downarrow}$ and $N_-=N_{\uparrow}-N_{\downarrow}=0$ due to the symmetry of the model. For $\lambda\rightarrow\infty$, $N_+$ has two BCS peaks at $\Delta\pm h$. As the domain wall becomes smaller, the inner peak is shifted towards the outer one and forms a shark-fin structure for a very small domain wall size [$\lambda=0.1\xi_0$ in Fig.\ref{fig:DensityOfStates}(a)]. In this case, although $N_-=0$, $N_+$ is different from the case with zero exchange field. Due to the spin-orbit relaxation like effect of the inhomogeneous exchange field, the peaks are wider and the superconducting gap is smaller compared to the case of $h=0$. 

Since the magnetization direction is opposite on the two sides of the domain wall, $N_s$ behaves differently in these two regions, such that $N_+(x)=N_+(-x)$ and $N_-(x)=-N_-(-x)$. In Fig.~\ref{fig:DensityOfStates}(b) we show $N_{\uparrow}$ at different location along the superconducting wire with a small domain wall $\lambda=0.1\xi_0$, in the absence of spin relaxation. These results in are similar with those in Ref.~\onlinecite{strambini2017revealing}, which concentrates on the limit $\lambda\ll\xi_0$. Here we further study the effect of spin relaxation on the local DoS in Fig.~\ref{fig:DensityOfStates}(c,d). The DoS in the presence of the two kinds of spin relaxation is plotted in Fig.~\ref{fig:DensityOfStates}(d) at the center of a domain wall with a size $\lambda=0.1\xi_0$. For comparison, the DoS at $x=-\infty$, which also refers to the homogeneous exchange field is plotted for the same parameters in Fig.~\ref{fig:DensityOfStates}(c). In both cases, spin-orbit relaxation broadens the peaks, but keeps $\Delta(x)$ unchanged. Spin-flip relaxation also broadens the peaks, but it suppresses $\Delta(x)$. 

These properties of the local DoS are caused by the inhomogeneous magnetization, and are needed for understanding the tunneling conductance as discussed in the next section. 

\section{Tunneling conductance\label{sec:Tunnelingcurrentandconductance}}
The local density of states is visible in measurements of a tunneling conductance through a barrier in contact with the FI/S bilayer. However, the results depend on whether the barrier itself is magnetic or not. Therefore, we consider either a non-magnetic tunnel contact (NISFI) or tunneling through the FI containing the domain wall (SFIN). In either system the tunneling current can be written as
$$
I=\frac{G_T}{e}\int_{-\infty}^{\infty}d\epsilon\bar{N}(\epsilon)\left[f(\epsilon-eV)-f(\epsilon)\right],
$$
where $G_T$ is the normal-state conductance, $V$ is the applied voltage and $f=\left[1+\exp(\epsilon/k_BT) \right]^{-1}$ is the Fermi function. Here the averaged density of states $\bar{N}(\epsilon)$ over the tunneling area is given by\cite{PhysRevB.86.214516}
$$
\bar{N}(\epsilon)=\frac{1}{L}\int_{x_a}^{x_b}dx\sum_{s=\uparrow,\downarrow}\left[1+sP(x)\right]N_s\left(\epsilon,x \right),
$$
where $P(x)$ is the spin polarization of the junction, $L=x_b-x_a$ is the size of the tunneling barrier, and $N_s$ is given in Eq.~\eqref{eq:DoSForSpinS}. Using the definition of $N_{\pm}=N_\uparrow\pm N_{\downarrow}$, we can also write
\begin{equation}\label{eq:AveDoS}
\bar{N}(\epsilon)=\frac{1}{L}\int_{x_a}^{x_b}dx\left[N_++P(x)N_- \right].
\end{equation}
The tunneling conductance is given by
\begin{equation}\label{eq:DefCond}
\frac{dI}{dV}(V)=\frac{G_T}{e}\frac{d}{dV}\int_{-\infty}^{\infty}d\epsilon\bar{N}(\epsilon)\left[f(\epsilon-eV)-f(\epsilon)\right].
\end{equation}

In the case of a scanning tunneling microscope measurement with a small normal metal tip ($x_b-x_a\ll\xi_0$ and $P(x)=0$), $\bar{N}=N_+$. In this case the tunneling conductance at $T\rightarrow0$ gives the local DoS $N_+$ shown in Fig.~\ref{fig:DensityOfStates}. 

\begin{figure}[t]
\includegraphics[width=1.0\linewidth]{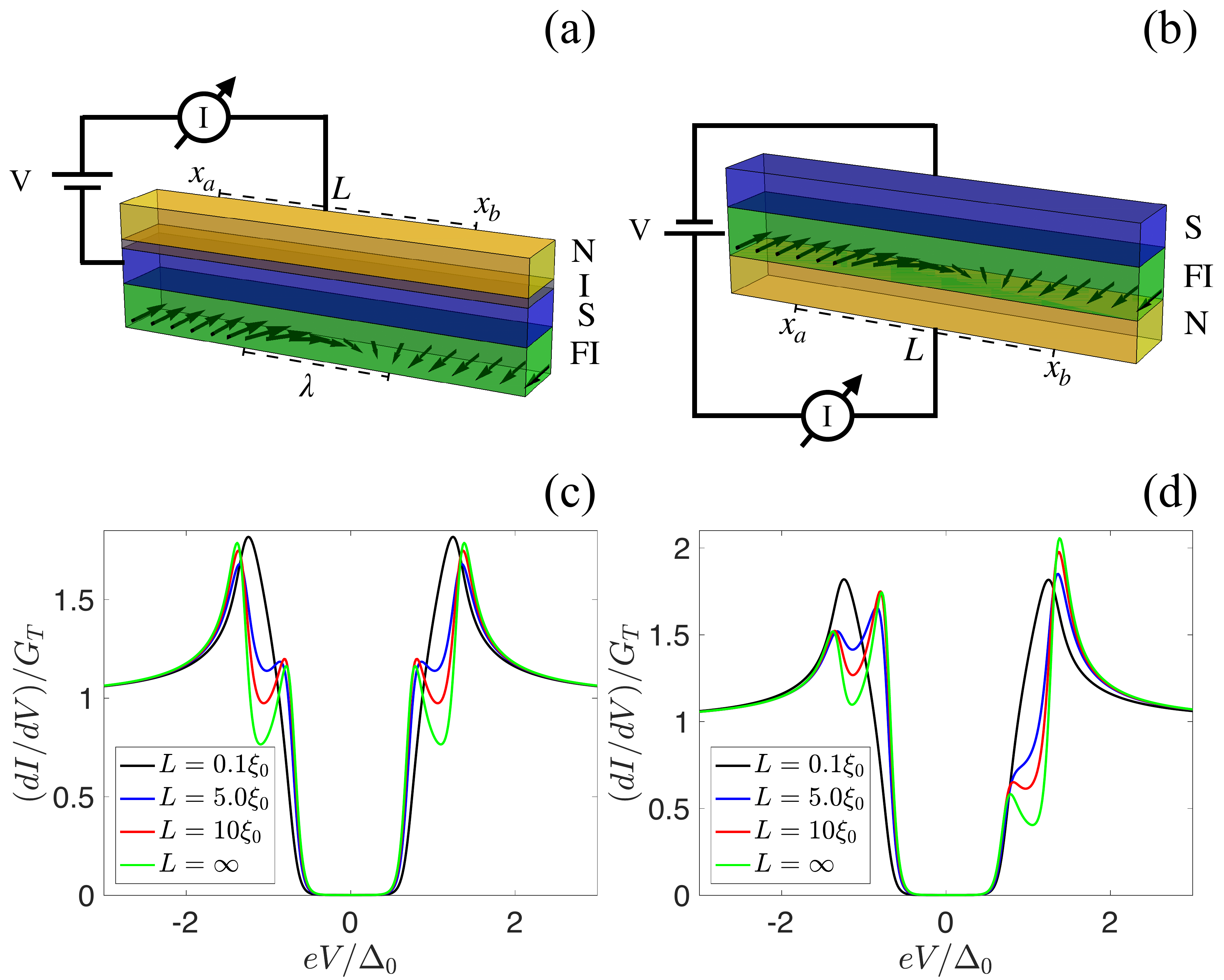}% Here is how to import EPS art
\caption{\label{fig:TunCorrCond} Tunneling conductance for different sizes of the tunneling barriers for NISFI and NISFI contacts, respectively. Here the center of the tunneling barrier is located at the center of the domain wall, $x_b=-x_a=L/2$. The calculations are carried out for a domain wall size $\lambda=0.1\xi_0$ at $T=0.05\Delta_0$ and $h=0.3\Delta_0$. In (d) the polarization is $P(x)=P_0\cos\alpha$, where $\alpha$ is the rotation angle, and $P_0=0.5$.}
\end{figure}

For a planar tunneling spectroscopy measurement with a normal metal electrode having size $x_b-x_a\geq\xi_0$, $\bar{N}$ is the averaged $N_+$ since $P(x)=0$. The tunneling conductance in this case is given in Fig.~\ref{fig:TunCorrCond}(c). We show the comparison of different sizes of tunneling barriers (centered at $x=0$). Since we are considering a single domain wall structure, a large tunneling barrier produces tunneling conductance identical with the case of homogeneous magnetization (green curve). If we choose a smaller tunneling barrier, the effect of the domain wall is more obvious. It reduces the effect of the exchange field, becomes similar to the case for a small homogeneous exchange field strength $h$, and reshapes the peaks in the tunneling conductance (red and blue curves).  

If the tunneling is through the ferromagnetic insulator, we need to include the effect of finite spin polarization. Assuming the polarization of the transmission to follow the local magnetization, the position dependent polarization $P$ is related to the rotation angle of the magnetization as $P(x)=P_0\cos\alpha$. The tunneling conductance is plotted in Fig.~\ref{fig:TunCorrCond}(d). The asymmetric dependence in the injection voltage in the conductance is a direct result of the polarization of the ferromagnet. Similar to Fig.\ref{fig:TunCorrCond}(c), for a large tunneling barrier the tunneling conductance is the same for the case of a homogeneous exchange field $h$ with a finite spin polarization $P_0$. The effect of the domain wall is again obvious for smaller tunneling barriers, and the spin-orbit relaxation like effect of the domain wall structure reduces the effect of the exchange field.

\section{Conclusion}
%\subsection{}
%\subsubsection{}
In conclusion, we have studied various properties of a superconductor in contact to a ferromagnet with a domain wall. We have studied the equilibrium spin current density, which exists due to the inhomogeneity of the exchange field. It also contributes to the superconducting free energy, which in turn affects the domain wall size. We show that the domain wall size is reduced by the contribution of the superconducting free energy to the domain wall energy. We have also studied the peculiar density of states around the domain wall, and the tunneling conductance. Our work can be a precursor to the study of the nonequilibrium effects, in particular domain wall motion.

\begin{acknowledgments}
We thank F. S. Bergeret, M. Silaev and S. Chakraborty for discussions. This project has received funding from the European Union’s Horizon 2020 research and innovation programme under grant agreement No. 800923 (SUPERTED) and the Academy of Finland projects No. 305256 and 317118.  
\end{acknowledgments}

\appendix

\section{Expression for the free energy}

\label{app:fsn}

For completeness, we give now a brief argument for the form of the
free energy in the superconducting state.  There are several ways
to arrive at such results, and below we follow a similar
procedure as in Ref.~\onlinecite{Eilenberger1968}.  The
free energy density difference between superconducting and normal
states can be written as (we set $N(0)=1$ in the following)
\cite{maki1964-pps1,gorkov59b}
\begin{align}
  \label{eq:coupling-constant}
  f_{sn}
  = -\int_0^{\gamma}\dd{\gamma'} \frac{\Delta\rvert_{\gamma'}^2}{(\gamma')^2}
  = \int_{\infty}^{1/\gamma}\dd{q} \Delta\rvert_{\gamma=1/q}^2
  \,.
\end{align}
The integral could now in principle be computed numerically from Usadel
equations, solving $\Delta$ self-consistently as a function of the coupling constant $\gamma$ in the geometry considered. This is cumbersome, and also unnecessary, as the integral can be evaluated analytically as follows.

Suppose there exists a functional $R=\int\dd{^3x}\,r$,
\begin{align}
  r[q,\Delta,\vec{z}] = q\Delta^2 + p[\Delta, \vec{z}]
  \,,
\end{align}
whose saddle point vs. $\vec{z}$ and $\Delta$
defines $\Delta(q)$. In other words, the 
self-consistency~\eqref{eq:selfconsg} and
Usadel equations~\eqref{eq:UsadelEquation} are defined by variations
\begin{align}
  \label{eq:variation-eqs}
  \frac{\delta R}{\delta\Delta}
  \rvert_* &= 0
  \,,
  &
  \frac{\delta R}{\delta z_j}
  \rvert_* &= 0
  \,,
\end{align}
and $\vec{z}$ is some parameterization of $\hat{g}$
under the constraint $\hat{g}^2=1$ of the Usadel equation.
Now,
\begin{align}
  \frac{d}{dq}R\rvert_*
  &=
  \partial_q R\rvert_*
  + \frac{\delta R}{\delta\Delta}\rvert_* * \frac{d\Delta_*}{dq}
  + \frac{\delta R}{\delta \vec{z}}\rvert_* * \frac{d\vec{z}_*}{dq}
  =
  \Delta_*^2
  \,,
\end{align}
where the last two terms vanish due to the saddle point conditions.
Therefore,
\begin{align}
  \label{eq:fnsR}
  f_{sn}
  =
  r[1/\gamma,\Delta_*,\vec{z}_*]
  -
  r[\infty,0,\vec{z}_{*,n}]
  \,.
\end{align}
This assumes there is a continuous solution branch connecting the
normal and the superconducting states as the coupling constant $\gamma$ (ie. $q$) is changed.

The next step is to find a suitable $R$ that satisfies
Eqs.~\eqref{eq:variation-eqs}. Its form can be guessed (or derived)
based on $\sigma$-model results \cite{belitz1994-amt,Altland1998,PhysRevB.64.014512}:
\begin{align}
  \label{eq:Rexpr}
  r
  &=
  q|\Delta|^2
  -
  2\pi{}T\sum_{\omega_n>0}
  \frac{1}{2}
  \tr\{
  -\frac{D}{4}(\nabla{}\check{g})^2
  +
  (\omega_n + i\mathbf{h}\cdot\boldsymbol{\sigma}) \hat{\tau}_3 \check{g}
  \\\notag&\qquad
  +
  (\Delta\hat{\tau}_+ + \Delta^*\hat{\tau}_-)\check{g}
  +
  \frac{1}{16\tau_{so}}(\boldsymbol{\sigma}\check{g})\cdot(\boldsymbol{\sigma}\check{g})
  \\\notag&\qquad
  +
  \frac{1}{16\tau_{sf}}(\boldsymbol{\sigma}\hat{\tau}_3\check{g})\cdot(\boldsymbol{\sigma}\hat{\tau}_3\check{g})
  \}
  \,.
\end{align}
First, we can note that variation vs $\Delta^*$ (or $\Delta$) gives
the self-consistency relation~\eqref{eq:selfconsg}.  For variations
with $\hat{g}$, we parametrize
$\delta\hat{g}=e^{\delta{}\hat{W}}\hat{g}e^{-\delta{}\hat{W}}-\hat{g} \simeq \delta W \hat{g} - \hat{g}
\delta \hat{W}$ to retain the normalization condition. Requiring variation
vs. $\hat{W}$ to vanish we find Eq.~\eqref{eq:UsadelEquation}.
Hence, the result has the property~\eqref{eq:variation-eqs}.

The sum defining $R$ is not convergent, and only the difference in Eq.~\eqref{eq:fnsR} is
well defined. There's also an implicit Matsubara frequency cutoff in the term that appears
in the self-consistency equation.  We can eliminate this issue by
substituting the self-consistency equation back into the $|\Delta|^2$ term in
Eq.~\eqref{eq:Rexpr}.  This results to Eq.~\eqref{eq:fsndef} in the
main text. Variation vs $\mathbf{h}$ still gives the correct quasiclassical expression
for the magnetization, but $\Delta$ and $\hat{g}$ can no longer be varied and have to be taken at their saddle-point values.

\bibliography{references}

\end{document}